\documentclass[12pt]{iopart}
\usepackage[dvips]{graphicx}

\begin{document}

\title[Existence of Dirac cones in diperiodic crystals]{Existence of Dirac cones in Brillouin zone of diperiodic atomic crystals according to group theory}

\author{V Damljanovi\'c and R Gaji\'c}

\address{Institute of Physics, University of Belgrade, Pregrevica 118, 11080 Belgrade, Serbia}
\ead{damlja@ipb.ac.rs}
\vspace{10pt}
\begin{indented}
\item[]October 2015
\end{indented}

\begin{abstract}
We have considered non-magnetic materials with weak spin-orbit coupling, that are periodic in two non-collinear directions, and finite in third, orthogonal direction. 
In some cases, combined time-reversal and crystal symmetry of such systems, allows the existence of Dirac cones at certain points in the reciprocal space. We have investigated in a systematic way, all points of Brillouin zone of all 80 diperiodic groups and have found sufficient conditions for the existence of $s=1/2$ Dirac fermions, with symmetry-provided band touching at the vertex of the Dirac cones.
Conversely, complete linear dispersion is forbidden for orbital wave-functions belonging to two-dimensional irreducible representations (irreps) of little groups that do not satisfy certain group-theoretical conditions given in this paper. Our results are illustrated by a tight-binding example.
\end{abstract}

%
\vspace{2pc}

\section{Introduction}

Dirac materials \cite{rev1,rev2,rev3} have physical properties that are well described by the effective Hamiltonian that resembles relativistic Quantum electrodynamics. Such materials include graphene, topological insulators, high temperature superconductors or Weyl semimetals. Interesting properties of these materials such as zero effective mass or high electron mobility, make them topics of intensive experimental and theoretical research. From the theoretical side, there have been attempts \cite{dpeng08,gra13} to predict new systems with Dirac-like dispersion. Especially interesting is the connection of the symmetry of the investigated systems with the properties of Dirac materials. Aoki and Shima \cite{ShAo93,AoSh94} have investigated lateral superstructures with a honeycomb symmetry within the tight-binding model. They have found that in some cases, the symmetry enforces the existence of dispersionless bands. Ma\~nes et al. \cite{Manj07} have determined the role that space-time inversion symmetry has in stabilizing the Fermi points in multilayer graphene. A generalized von Neumann-Wigner theorem was formulated by Asano and Hotta to give number of constrains on the lattice for the existence of an accidental energy band contacts with linear dispersions. Some of the constrains are ensured by inversion, reflection or time-reversal symmetries \cite{AsHo11}. Recently, rotational symmetries are used to classify three-dimensional Dirac semimetals (3D analogue of graphene) with accidental band crossings \cite{NatCom14}. Earlier, Young et al. showed that physics of graphene can be extended to three-dimensional materials. They used theory of double space groups to examine existence of Dirac points in 3D crystals belonging to the space group of diamond or zinc blende lattice \cite{YoZa12}. At about the same time, by skilful use of space group representation theory, J. L. Ma\~nes have investigated points of symmetries in the Brillouin zone of all 230 space groups, and have found that the crystal symmetry combined with the time-reversal symmetry (TRS), leads to the existence of the Weyl points in systems belonging to a limited number of space groups \cite{Manj12}.

In this paper we have considered non-magnetic, single-crystalline materials with weak spin-orbit coupling, that are periodic in two directions, but finite in third, orthogonal direction. Such atomic systems must belong to one of the 80 diperiodic groups \cite{itce}. We have investigated all points of Brillouin zone of all 80 diperiodic groups and have found sufficient conditions for the existence of $s=1/2$ Dirac fermions. Symmetry also provides the band touching at the vertex of the cones. In spite of the extensive research related to the interplay between the symmetry and the properties of Dirac (Weyl) materials, to best of our knowledge no systematic analysis of the connection between symmetry of diperiodic systems and the appearance of Dirac cones in their energy spectrum were performed until now. Our paper therefore aims to fill in this gap. Apart from the general idea, the results of \cite{Manj12} can not be just simply applied to our case, where the reduced dimensionality of the system leads to different mathematical conditions. For example, a combined time-reversal and crystal symmetry provides the Dirac cones at the Brillouin zone corners of eight diperiodic groups belonging to the hexagonal system. At the end of the paper, we provided an example to illustrate our results.

\section{Method}

Let $\bi{k_{0}}$ be a real vector from (two-dimensional) Brillouin zone of a diperiodic group and $\bi{q}$ a two-dimensional, real vector of small modulus. $G(\bi{k_{0}})$ is a little group of the wave vector (a set of all elements of diperiodic group which rotational parts transform $\bi{k_{0}}$ to it's equivalent) and $G_{0}(\bi{k_{0}})$ is a point group of the wave vector (a set that consists of rotational parts of elements of $G(\bi{k_{0}})$ only). Further, let $e_j(\bi{r},\bi{k_0})$, be orbital wave functions belonging to an irreducible representation (irrep) $R$ of $G(\bi{k_{0}})$. For any diperiodic group such irreps are either one- or two-dimensional \cite{gdg}. Since $e_j(\bi{r},\bi{k_0})$ are Bloch functions, $R$ must be allowed \cite{bcs} (relevant \cite{Crnw}, small \cite{BrCr}) irrep of $G(\bi{k_{0}})$. Let $[\hat{H}(\bi{k_0})]_{jl}=\left\langle e_j(\bi{r},\bi{k_0})\right|\hat{H}_0(\bi{r})\left|e_l(\bi{r},\bi{k_0})\right\rangle$. The original Hamiltonian $\hat{H}_0(\bi{r})$ is real. Lets assume that $R$ is two dimensional. It follows, for every element $(\hat{g}\left|\bi{t}\right)$ (in Seitz notation) of $G(\bi{k_{0}})$:
\begin{equation}
  e_j((\hat{g}\left|\bi{t}\right)^{-1}\bi{r},\bi{k_0})=\sum_{l=1}^2[\hat{R}(\hat{g}\left|\bi{t}\right)]_{lj}e_l(\bi{r},\bi{k_0}).
\end{equation}
The Taylor expansion of the Hamiltonian $\hat{H}(\bi{k})$ in the vicinity of the point $\bi{k_{0}}$ reads \cite{KoKa07,MoKa10}:
\begin{equation}
\hat{H}(\bi{k_0}+\bi{q})\approx\hat{H}(\bi{k_0})+\sum_{j=0}^3(\bi{v}_{j}\cdot\bi{q})\hat{\sigma}_j,
\end{equation}
where $\hat{\sigma}_{1,2,3}$ ($\hat{\sigma}_0$) are usual Pauli matrices (is two-dimensional unit matrix) and
\begin{equation}
\left(\forall j=0,1,2,3\right)\ \bi{v}_{j}=\frac{1}{2}\left\{\frac{\partial}{\partial\bi{q}}Tr[\hat{\sigma}_{j}\hat{H}(\bi{k_0}+\bi{q})]\right\}_{\bi{q}=0},
\end{equation}
are two-dimensional, real vectors. If we assume double degeneracy of the electron energy level $E_0$ at $\bi{k_0}$, we get for the energy in the vicinity of this point:
\begin{eqnarray}
\label{4}
E_{1,2} & = & E_0+\bi{q}\cdot\bi{v}_0\pm\sqrt{\sum_{j=1}^3(\bi{q}\cdot\bi{v}_j)^2}\nonumber\\ & = & E_0+\bi{q}\cdot\bi{v}_0\pm\sqrt{u_1 q_1^2+u_2 q_2^2},
\end{eqnarray}
where
\begin{equation}
u_{1,2}=\frac{1}{2}\sum_{j=1}^3\bi{v}_j^2 \pm\frac{1}{2}\sqrt{\left(\sum_{j=1}^3\bi{v}_j^2\right)^2-4\sum_{j=1}^2\sum_{l=j+1}^3(\bi{v}_j\times\bi{v}_l)^2},
\end{equation}
are eigenvalues of the quadratic, two-dimensional matrix $\hat{S}=\sum_{j=1}^3 \left|\bi{v}_j\right\rangle\left\langle \bi{v}_j\right|$, $q_{1,2}$ are projections of $\bi{q}$ along orthonormalized eigenvectors of $\hat{S}$, $\left\langle \bi{v}_j\right|^T=\left|\bi{v}_j\right\rangle=\bi{v}_j$ and $T$ denotes transposition.  The matrix $\hat{S}$ is symmetric and positively semi-definite, so $u_1 \geq u_2 \geq 0$.
For $u_{1,2}$ both different than zero, (\ref{4}) represents a pair of Dirac cones that is tilted for non-zero $\bi{v}_0$. Combined TRS and crystal-symmetry can make one or both $u$'s vanish. In what follows we will investigate means to avoid such cases.

\section{Results}

By the Taylor expansion of corresponding commutation relations between the Hamiltonian $\hat{H}(\bi{k})$ and matrices of irreps of $G(\bi{k})$ around $\bi{k}_0$ up to first order \cite{Manj12}, we get
\begin{equation}
\label{eqW}
\hat{W}=\hat{R}^*\left((\hat{g}\left|\bi{t}\right)\right)\hat{W}\left(\hat{R}^T\left((\hat{g}\left|\bi{t}\right)\right)\otimes\hat{g}'^T\right),
\end{equation}
where $(\hat{g}\left|\bi{t}\right)$ is an element of $G(\bi{k_{0}})$, $\hat{g}'$ is the reduction of $\hat{g}$ to the diperiodic plane, $\hat{R}\left((\hat{g}\left|\bi{t}\right)\right)$ is matrix of the irrep $R$ of $G(\bi{k_{0}})$ that corresponds to the element $(\hat{g}\left|\bi{t}\right)$, $*$ is complex conjugation, $\otimes$ denotes the Kronecker product and
\begin{equation}
\hat{W}^*=\sum_{j=0}^3\hat{\sigma}_j\otimes\left\langle \bi{v}_j\right|.
\end{equation}
We will now state group-theoretical conditions that ensure the existence of Dirac cones touching at vertex.

$O_1$: if $R$ is two-dimensional irrep of $G(\bi{k_{0}})$,  two bands will touch at $\bi{k}_0$. 

$O_2$: in order to make bands split along every direction away from the $\bi{k_{0}}$, this point must be of locally maximal symmetry. 

$O_3$: in order $\bi{v}_1,\;\bi{v}_2,\;\bi{v}_3$ not to vanish simultaneously, $(R^*\otimes R-\Gamma_1)\otimes\Gamma_{2DPV}$ must contain at least one $\Gamma_1$, where $\Gamma_{2DPV}$ is the two-dimensional polar-vector representation of $G_0(\bi{k_{0}})$ and $\Gamma_1$ is the totally symmetric (unit) representation of $G_0(\bi{k_{0}})$. 

$O_4$: in order $u_2$ not to vanish from symmetry reasons, in addition to $O_3$, $(R^*\otimes R-\Gamma_1)\otimes det\left(\Gamma_{2DPV}\right)$ must contain at least one $\Gamma_1$, where the number corresponding to an element $\hat{g}$ of $G_0(\bi{k_{0}})$ by an irrep  $det\left(\Gamma_{2DPV}\right)$ is $det(\hat{g}')$. 

Note that since $R^*\otimes R$ always contains $\Gamma_1$, the conditions $O_3$ and $O_4$ are mathematically well-defined. These two conditions are derived from (\ref{eqW}). The next conditions are due to TRS (reality of the original Hamiltonian $\hat{H}_0(\bi{r})$).

$O_5$: for $-\bi{k_0}$ equivalent to $\bi{k_0}$, if $R$ is a real irrep then $u_2=0$, but if $R$ is pseudo-real or complex, then the energy level $E_0$ is four times degenerate and the presence of Dirac cones must be investigated case-by-case.

$O_6$: if $-\bi{k_0}$ is not equivalent to $\bi{k_0}$ but there exist an element $(\hat{h}\left|\bi{t}\right)$ of the diperiodic group such that $\hat{h}\bi{k_0}$ is equivalent to $-\bi{k_0}$ then combination of this element and complex conjugation leads in principle to an additional constrain \cite{Manj12}. However, it turned out that when conditions $O_1$-$O_4$ were fulfilled, the condition $O_6$ led to no further restrictions.

Finally, we state the condition that shows when the Dirac cones predicted to exist by $O_1$-$O_6$ are tilted. In order to make our results more general, we didn't use this condition.

$O_7$: for Dirac cones to be untilted ($\bi{v}_0=0$), $\Gamma_{2DPV}$ must not contain $\Gamma_1$.

We have used conditions $O_1$-$O_6$ to investigate Brillouin zones \cite{gdg,Dzam98} of all eighty diperiodic groups. It turned out that only eight diperiodic groups fulfilled the conditions. All of them are symmorphic and belong to the hexagonal system, so their Brillouin zone is hexagon. Symmetry allowed, isotropic Dirac cones are located at the corners of hexagons (K-points) and groups are listed in the table \ref{tab1}. 

\begin{table}
\caption{\label{tab1} Diperiodic groups having symmetry-allowed Dirac cones at the Brillouin zone corners (K-points). All groups are semi-direct product of the translational subgroup $\overline{T}$, and a point group. Notations of diperiodic groups are according to \cite{itce}. The responsible irrep denotes irrep that generates a pair of Dirac cones. The point group of the wave vector is given in parentheses.}
\begin{indented}
\item[]\begin{tabular}{@{}llll}
\br
\multicolumn{3}{l}{Diperiodic group} & Responsible irrep ($G_0(K)$)\\
\mr
68 & $p\,3\,2\,1$ & $\overline{T}D_3$ & $E(D_3)$\\
70 & $p\,3\,1\,m$ & $\overline{T}C_{3v}$ & $E(C_{3v})$\\
71 & $p\,\overline{3}\,1\,2/m$ & $\overline{T}D_{3d}$ & $E(C_{3v})$\\
72 & $p\,\overline{3}\,2/m\,1$ & $\overline{T}D_{3d}$ & $E(D_3)$\\
76 & $p\,6\,2\,2$ & $\overline{T}D_6$ & $E(D_3)$\\
77 & $p\,6\,m\,m$ & $\overline{T}C_{6v}$ & $E(C_{3v})$\\
79 & $p\,\overline{6}\,2\,m$ & $\overline{T}D_{3h}$ &  $E'(D_{3h})$, $E"(D_{3h})$\\
80 & $p\,6/mmm$ & $\overline{T}D_{6h}$ & $E'(D_{3h})$, $E"(D_{3h})$\\
\br
\end{tabular}
\end{indented}
\end{table}

\section{Discussion}

\begin{figure}
\includegraphics[width=\columnwidth]{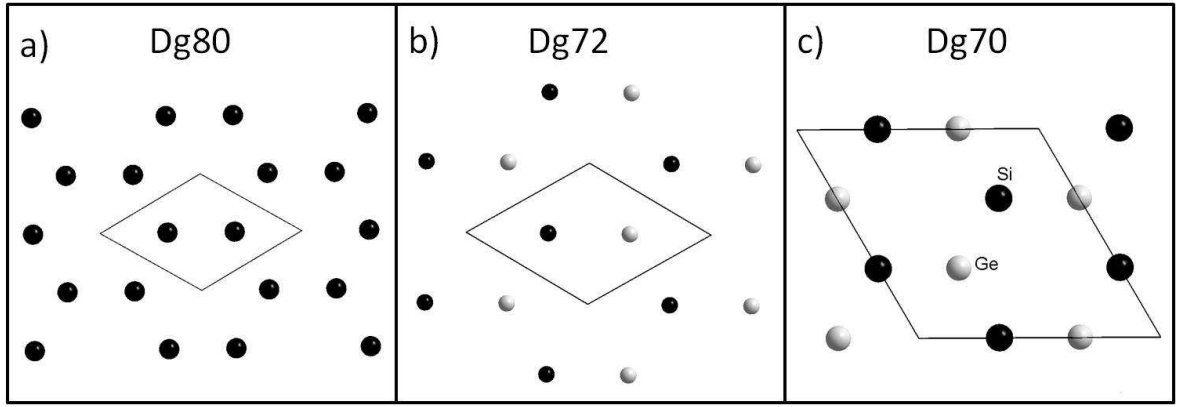}
\caption{\label{slka1} Top view of the crystal structure for a) germanene (after \cite{SiGeplnr}), b) low-buckled silicene (after \cite{SiGepckrd}) and c) silicon germanide monolayer (after \cite{SilGer}). Dark (light) balls represent nuclei that are located above (below) the drawing plane. The black rhombus represents the primitive unit cell. The notations for diperiodic groups are according to \cite{itce}.}
\end{figure}

\begin{figure}
\includegraphics[width=\columnwidth]{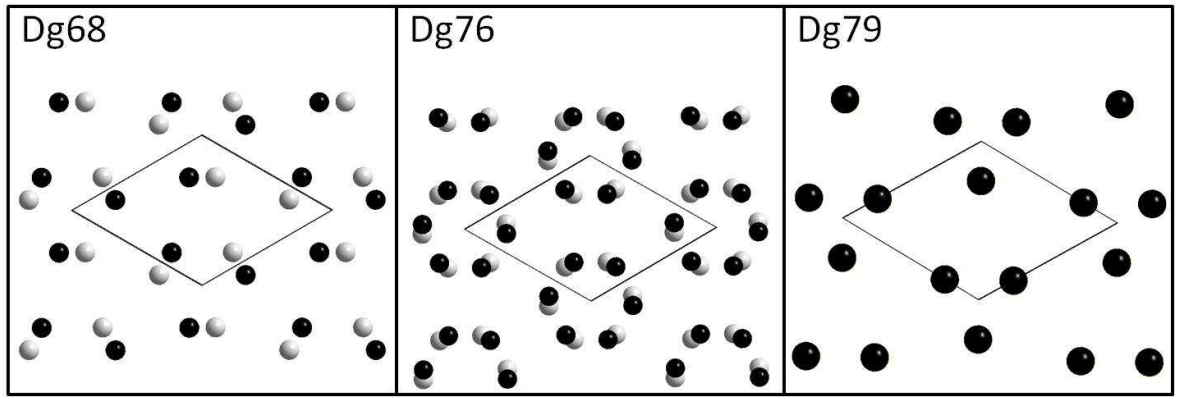}
\caption{\label{slk2} Examples of crystal structures for some groups from the table \ref{tab1} (top view). All nuclei are of the same type. Dark (light) balls represent nuclei that are located above (below) the drawing plane. For the group Dg79 light balls are exactly below the dark ones. The black rhombus represents the primitive unit cell. The notations for diperiodic groups are according to \cite{itce}.}
\end{figure}

We can see that the table \ref{tab1} contains Dg80 - the symmetry group of mono-layer graphene. The same group is the symmetry group of the kagome lattice, which has Dirac cones at K-points, within a tight-binding model \cite{kago10}. Planar, graphene-like silicon (silicene) and germanium (germanene) also have Dirac cones at the K-point, within tight-binding \cite{SiliTB07} and \emph{ab initio} \cite{SiGeplnr} method. In addition, \emph{ab initio} calculations on non-planar, low-buckled silicene and germanene, show Dirac cones in energy dispersion \cite{SiGepckrd}. This configuration belongs to the diperiodic group Dg72 that is listed in the table \ref{tab1}. Recently, the first principles calculations have shown a pair of Dirac cones at the $K$-point of the low-buckled $SiGe$ compound \cite{SilGer}. This compound belongs to the group Dg70, listed in the table \ref{tab1}. The crystal structures of these compounds are shown in the figure \ref{slka1}.
On the other hand, the mono-layer $MoS_2$ belongs to Dg78, that does not satisfy our conditions. Dirac cones in the energy spectrum of this material are absent \cite{MoS2knjg}. 
Examples of crystal structures belonging to other groups from the table \ref{tab1} are given in the figure \ref{slk2}.
All shown structures generate Dirac cones regardless of the type of orbitals used for building the wave functions. For example, the structure presented in figure \ref{slk2} belonging to Dg68, would have two pairs of Dirac cones within the tight-binding model from the $s$-type orbitals, if one takes enough neighbors into account. For completness, one has to note that the figure \ref{slk2} does not exhaust all possibilities. It is also important to note that it is not guaranteed that the Fermi level will cross the energy near the contact points of the cones.
Note that in order Dirac cones to appear, orbital wave-functions must transform according to a responsible irrep from the table \ref{tab1}. Orbital double degeneracy can appear also when two mutually conjugated, one-dimensional irreps related by TRS, form two-dimensional physically irreducible representation. Since then $u_2=0$, these cases were excluded from the consideration.
The condition $O_1$ is sufficient (but not necessary) for the band touching at $\bi{k_0}$. For this reason Dirac cones touching each other accidentally, can appear in other parts of Brillouin zone and in other diperiodic groups, then listed in the table \ref{tab1}. Such is the case, for example, in $S$-graphene, which belongs to the rectangular system and whose anisotropic Dirac cones appear at points belonging to $\Gamma Y$ and $MX$ directions of it's Brillouin zone \cite{Sgphn}. Alternatively, in systems where orbital and spin degrees of freedom can not be separated, above analysis does not apply. Some topological insulators, where the spin-orbit coupling is large, have a pair of Dirac cones at the Brillouin zone center (the $\Gamma$ -point) \cite{topiz}.

\section{An illustrative example}
\begin{figure}
\includegraphics[width=\columnwidth]{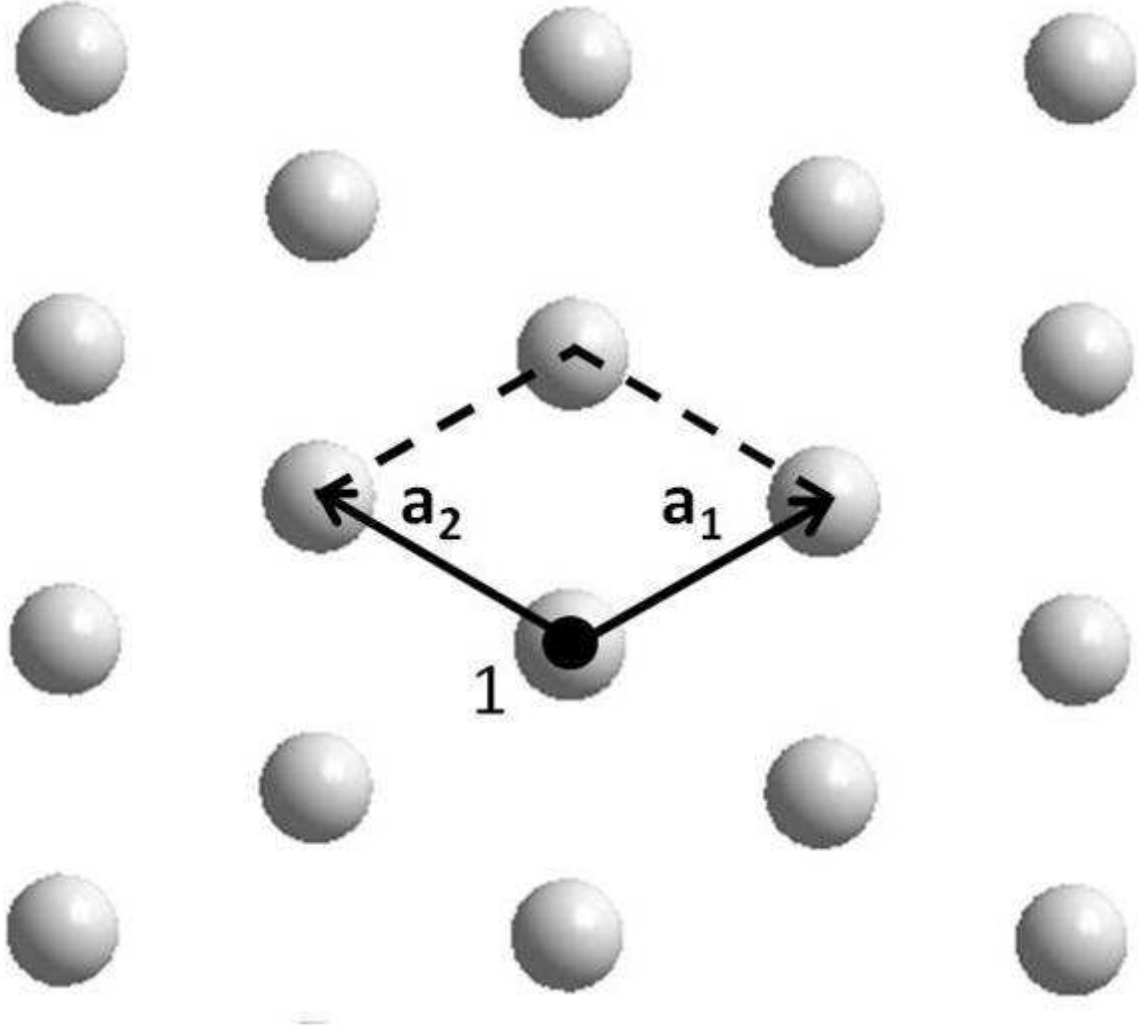}
\caption{\label{slika3} The crystal structure of the example with the adopted primitive unit cell.}
\end{figure}

As an example we will consider a tight-binding model on a structure that arises from the site 1a of diperiodic group Dg80. The crystal structure of the example is shown in the figure \ref{slika3}.
In order to ensure presence of at least one pair of orbital wave-functions that belong to a responsible irrep from the table \ref{tab1}, we can choose for this particular example any type of orbitals except the $s$-type. For the simplicity let's use $p$-orbitals for building the tight-binding wave-functions. The primitive cell of this structure contains one atom which is located at the origin. Let $\bi{a_1}$ and $\bi{a_2}$ be primitive lattice  vectors as indicated in the figure \ref{slika3}, the orbital $f_1=(\sqrt{3}p_x+p_y)/2$ is directed along $\bi{a_1}$ and $f_2=p_y$ is directed along $\bi{a_1}+\bi{a_2}$. $f_3=p_z$ where the z-axis is perpendicular to the diperiodic plane. Electronic levels of such a system are classified as $A''_2+E'$ for the K-point (point group $D_{3h}$). In contrast to graphene, which $p_z$ orbitals located at two atoms in the primitive cell belong to the irrep $E''$, the in-plane orbitals of our example belong to the irrep $E'$. Both irreps generate a pair of Dirac cones. Six nearest neighbors of the central atom are located at vertices of regular hexagon, with the central atom located at it's center. By taking into account the system's symmetry, we get for it's Hamiltonian:

\begin{eqnarray}
\left[\hat{H}(\bi{k})\right]_{11}=t_2+t_3w+(t_4-t_3)r_1, \nonumber \\
\left[\hat{H}(\bi{k})\right]_{22}=t_2+t_3w+(t_4-t_3)r_3, \nonumber \\
\left[\hat{H}(\bi{k})\right]_{12}=\left[\hat{H}(\bi{k})\right]_{21}=
\frac{1}{2}t_2+\frac{1}{2}t_4w+(t_3-t_4)r_2,\\
\left[\hat{H}(\bi{k})\right]_{13}=\left[\hat{H}(\bi{k})\right]_{31}
=\left[\hat{H}(\bi{k})\right]_{23}=\left[\hat{H}(\bi{k})\right]_{32}=0, \nonumber \\
\left[\hat{H}(\bi{k})\right]_{33}=t_0+t_1w, \nonumber
\end{eqnarray}
where

\begin{eqnarray}
\label{eq9}
t_0=\frac{1}{\sqrt{N}}\int f_3(\bi{r})\hat{H}_0(\bi{r})f_3(\bi{r})d^3\bi{r}, \nonumber \\
t_1=\frac{2}{\sqrt{N}}\int f_3(\bi{r})\hat{H}_0(\bi{r})f_3(\bi{r}-\bi{a_1})d^3\bi{r}, \nonumber \\
t_2=\frac{1}{\sqrt{N}}\int f_2(\bi{r})\hat{H}_0(\bi{r})f_2(\bi{r})d^3\bi{r}, \\
t_3=\frac{2}{\sqrt{N}}\int f_2(\bi{r})\hat{H}_0(\bi{r})f_2(\bi{r}-\bi{a_1})d^3\bi{r}, \nonumber \\
t_4=\frac{2}{\sqrt{N}}\int f_2(\bi{r})\hat{H}_0(\bi{r})f_2(\bi{r}-\bi{a_1}-\bi{a_2})d^3\bi{r}, \nonumber
\end{eqnarray}
and $N$ is the number of primitive cells in the system. In (\ref{eq9}) the integration is over the whole real, three-dimensional space. In addition:
\begin{eqnarray}
r_1=\cos(\bi{k}\cdot\bi{a_1}), \nonumber \\
r_2=\cos(\bi{k}\cdot\bi{a_2}), \nonumber \\
r_3=\cos[\bi{k}\cdot(\bi{a_1}+\bi{a_2})], \nonumber \\
w=\sum_{j=1}^3 r_j. \nonumber
\end{eqnarray}
To obtain energy levels we have to solve the following equation for $\epsilon$:
\begin{equation}
det\left[\hat{H}(\bi{k})-\epsilon \hat{S}\right]=0,
\end{equation}
where 
\begin{equation}
\hat{S}=\left[
\begin{array}{ccc}
	1 & 1/2 & 0 \\
	1/2 & 1 & 0 \\
	0 & 0 & 1 \\
\end{array}
\right],
\end{equation}
is taking into account the fact that starting wave functions are not all mutually orthogonal. The solution of this problem is:
\begin{eqnarray}
\epsilon_{1,2} & = & t_2+\frac{2}{3}(t_3+\frac{1}{2}t_4)w\pm \nonumber \\
 & \pm & \frac{2}{3}\left|t_4-t_3\right|\sqrt{\sum_{j=1}^3 r_j^2-\sum_{3\geq l>j\geq 1}r_jr_l}, \\
\epsilon_3 & = & t_0+t_1w. \nonumber
\end{eqnarray}
We take $\bi{k}=(\bi{b_1}+\bi{b_2})/3+\bi{q}$, the expansion around the K-point. $\bi{b_1}$, $\bi{b_2}$ are primitive vectors of the reciprocal lattice such that $\bi{b_j}\cdot\bi{a_l}=2\pi\delta_{j,l}$. For small $q=\left|\bi{q}\right|$ we get ($a=\left|\bi{a_1}\right|=\left|\bi{a_2}\right|$):
\begin{equation}
\epsilon_{1,2}\approx t_2-(t_3+\frac{1}{2}t_4)\pm\frac{\sqrt{3}}{2}a\left|t_4-t_3\right|q,
\end{equation}
which, since $t_4\neq t_3$, presents a pair of Dirac cones, exactly as predicted by our theory!

For the $\Gamma$-point (the center of the Brillouin zone), the in-plane orbitals belong to an irrep $E_{1u}$ of the point group $D_{6h}$. This irrep does not satisfy the condition $O_3$ so the dispersion in the vicinity of this point should be quadratic. The expansion around $\bi{k_0}=0$, for small $q$ reads:
\begin{equation}
\epsilon_{1,2}\approx t_2+2t_3+t_4-\frac{1}{2}a^2\left(t_3+\frac{1}{2}t_4\pm\frac{1}{2}\left|t_3-t_4\right|\right)q^2.
\end{equation}
The quadratic dispersion in the last formula is another confirmation of our prediction.

\section{Conclusions}

In summary, we have determined the points in Brillouin zone of non-magnetic, diperiodic atomic crystals, with weak spin-orbit interaction, where symmetry allows linear, cone-like dispersion in the vicinity of these points. We have formulated a set of group-theoretical conditions which guarantee Dirac-like energy dispersion. Out of all eighty diperiodic groups investigated, only eight of them fulfill these conditions. All of them are symmorphic and belong to hexagonal system. For all of them, Dirac cones are located at the corners of hexagons (K-points) that present borders of the Brillouin zone. Our prediction is confirmed by tight-binding and density-functional theory calculations on numerous examples published in the literature. On the other hand, Dirac-type dispersion is symmetry-forbidden for orbital wave-functions belonging to any allowed, two-dimensional irrep that is not listed in the table \ref{tab1}. In the course of work, we have also investigated all other parts of the Brillouin zone, i.e. without the condition $O_2$. We have found that for some non-symmorphic diperiodic groups, there are lines of symmetry in the Brillouin zone which fulfill the conditions $O_1$ and $O_3$-$O_6$. In such cases, there is only Dirac-like dispersion in the direction that is perpendicular to those lines. This is another example where the absence of complete Dirac cones is caused by the \emph{presence of to many band contacts} \cite{Wang13}.

One of the conditions used, is sufficient but not necessary, so our analysis does not exhaust all cases. Two bands can touch each other accidentally but even in that case, there are group-theoretical conditions which must be satisfied in order to avoid quadratic dispersion. An analysis of these cases are beyond of the scope of this paper, as well as the question of stability of Dirac cones, or how to achieve that the Fermi level crosses the energy at the contact point of the cones.

\ack{This work was supported by the Serbian Ministry of Education, Science and Technological Development under project numbers OI 171005 and III 45016.}

\section*{References}


\end{document}